\documentclass[modern]{aastex63}
\usepackage{amsmath,amssymb,bm}
\def \Win {W_{\rm in}}
\def \Wres {W_{\rm res}}
\def \Wout {W_{\rm out}}
\def \Wfb {W_{\rm fb}}

\def \xx {{\bm x}}
\def \yy {{\bm y}}
\def \yhat {\hat{{\bm y}}}

\def \del2z {\partial^{2}_{z}}

\def \A {{\bm A}}

\def \f {{\bm f}}

\def \bb {{\bm b}}

\def \B {{\bm B}}

\def \k {{\bm k}}
\def \x {{\bm x}}

\def \t  {\theta}
\def \p  {\phi}

{}

\def \B {\bm B}

\def \A {\bm A}

 \def \a {{\alpha_0}}

\renewcommand{\fig}[1]{Fig.~\ref{#1}}

\def\drawing #1 #2 #3 {
\begin{center}
\setlength{\unitlength}{1mm}
\begin{picture}(#1,#2)(0,0)
\put(0,0){\framebox(#1,#2){#3}}
\end{picture}
\end{center} }

\usepackage{etoolbox}
\shorttitle{Forecasting Solar Cycle 25}
\shortauthors{A.Espuña-Fontcuberta et al.}
\graphicspath{{./}{figures/}}

\begin{document}
\title{A model-free, data-based forecast for sunspot cycle 25}
\correspondingauthor{Dhrubaditya Mitra}
\email{dhruba.mitra@gmail.com}
\author{Aleix Espuña-Fontcuberta }
\affiliation{NORDITA, Royal Institute of Technology and 
Stockholm University, Roslagstullsbacken 23, SE-10691 Stockholm, Sweden}
\author[0000-0003-2638-6047]{Saikat Chatterjee}
\affiliation{School of Electrical Engg and Computer Science, KTH Royal Institute of Technology, Sweden}
\author[0000-0003-4861-8152]{Dhrubaditya Mitra}
\affiliation{NORDITA, Royal Institute of Technology and Stockholm University, 
Roslagstullsbacken 23, SE-10691 Stockholm, Sweden}
\author[0000-0001-5205-2302]{Dibyendu Nandy}
\affiliation{Center of Excellence in Space Sciences India and Department of Physical Sciences, Indian Insitute of Science Education and Research Kolkata, Mohanpur 741246, India}

 \begin{abstract}
    The dynamic activity of the Sun, governed by its cycle of sunspots -- strongly magnetized regions that are observed on its surface -- modulate our solar system space environment creating space weather. Severe space weather leads to disruptions in satellite operations, telecommunications, electric power grids and air-traffic on polar routes. Forecasting the cycle of sunspots, however, has remained a challenging problem. We use reservoir computing -- a model-free, neural--network based machine-learning technique -- to  forecast the upcoming solar cycle, sunspot cycle 25. The standard algorithm forecasts that solar cycle 25 is going to last about ten years, the maxima is going to appear in the year 2024 and the maximum number of sunspots is going to be 113 ($\pm15$). We also develop a novel variation of the standard algorithm whose forecasts for duration and peak timing matches that of the standard algorithm, but whose peak amplitude forecast is 124 ($\pm2$) -- within the upper bound of the standard reservoir computing algorithm. We conclude that sunspot cycle 25 is likely to be a weak, lower than average solar cycle, somewhat similar in strength to sunspot cycle 24.  
\end{abstract}
\section*{}

The Sun shows magnetic activity over a large range of length and time scales. 
The total number of sunspots seen in the Sun varies with an 
approximately 11-year cycle. 
This cycle itself is not a regular one, its amplitude varies with time with no particular regularity~\citep{hathaway2015solar}.
The solar cycle plays an important role in governing space weather
which in turn has a major impact on our modern society. These include disruptions of satellite operations that impacts telecommunication 
networks and global positioning systems, geomagnetic storms that lead to electric power grid failures and air-traffic disruptions over polar routes~\citep{schrijver2015understanding}. The economic cost of a severe magnetic storm, say, e.g., of the magnitude of the great magnetic storm of 1859 -- the Carrington event -- is estimated to be greater than the economic cost of hurricane 
Katrina~\citep{space_weather_report_usacademy08}. 

Such considerations have spurred the field of solar cycle forecasting with diverse techniques employed to forecast upcoming solar cycles. \cite{petrovay2010solar} classified such techniques in to three groups: model--based methods, precursor methods and extrapolation methods.  Each has their strengths and weaknesses. Most importantly, the first two are closely connected with the physical insight of the solar dynamo that determines the solar cycle and last one is model agnostic and data--based. The solar magnetic cycle is thought to originate in a dynamo mechanism through complex, non-linear interactions between magnetic fields and plasma flows in the Sun's convection zone~\citep{charbonneau2010LRSP}. The extreme conditions and turbulent nature of stellar convection zones, combined with a lack of observational constraints, make computational modeling of the solar dynamo mechanism quite challenging.There have been only a few model-based forecast for the sunspot cycle which has just concluded, cycle 24~\citep{dikpati2006predicting,dikpati2006simulating, dikpati2007simulating,choudhuri2007predicting,jiang2007solar}. However, these model-based forecasts were highly inconsistent -- which was a result of differing assumptions regarding the turbulent nature of the Sun's convection zone~\citep{yeates2008exploring}. A NASA-NOAA Panel that typically attempts to generate a consensus prediction before the start of a sunspot cycle made an early forecast of a very strong solar cycle 24 which proved to be incorrect. In fact, this panel revised their forecast to a weak cycle 24 a few years following their first forecast. This indicates the uncertainty and challenges in predicting solar cycles.    

We note that terrestrial weather forecasting follows a similar route, although it is a relatively more mature field. Moreover, the solar dynamo model and its parameters are not as well constrained by observations as models for terrestrial weather forecasting.  

A recently developed physical model based forecasting scheme relied on coupling two distinct models of magnetic flux transport on the Sun, namely a solar surface flux transport model and a solar convection zone dynamo model~\citep{bhowmik2018prediction}. This physics-based modeling technique was quite successful in hind-casting and matching nearly a century of solar cycle observations and predicted a weak, but not insignificant solar cycle 25 similar or slightly stronger to the previous cycle peaking in 2024. Given major advances in both understanding the theory of solar cycle predictability, as well as application of data-based machine learning techniques to solar cycle forecasting, it would be interesting to assess at this juncture whether the best of these very diverse techniques result in predictions that are consistent with each other.

The last decade has found machine-learning techniques to be extraordinarily successful in making forecasts. They are particularly useful in those cases where a mechanistic model is either unavailable or poorly constrained, as is the case in many astrophysical and geophysical problems. They have played an increasingly important role in making data-based forecasts in several problems in solar physics~\citep{bobra2015solar, bobra2016predicting, dhuri2019machine}. Starting with \cite{fessant1996prediction} different neural networks have been used with varying degrees of success to forecast solar cycles~\citep{pesnell2012solar},
including few recent ones~\citep{pala2019forecasting,covas2019neural,benson2020forecast} who made forecasts for the upcoming cycle 25.

We use a particular technique called reservoir computing~\citep{jaeger2001echo,maass2002real,jaeger2004harnessing,lukovsevivcius2009reservoir} that has been used successfully to forecast  delay differential equations,  low-dimensional chaotic systems  and
even large spatiotemporally chaotic systems~\citep{pathak2018model}. This methods has so far not been used to forecast the solar cycle although theoretical considerations suggest that the solar dynamo mechanism can be represented by a system of delay differential equations~\citep{wilmotsmith2006timedelay}.

\begin{figure*}
\begin{center}
  \includegraphics[width=0.8\linewidth]{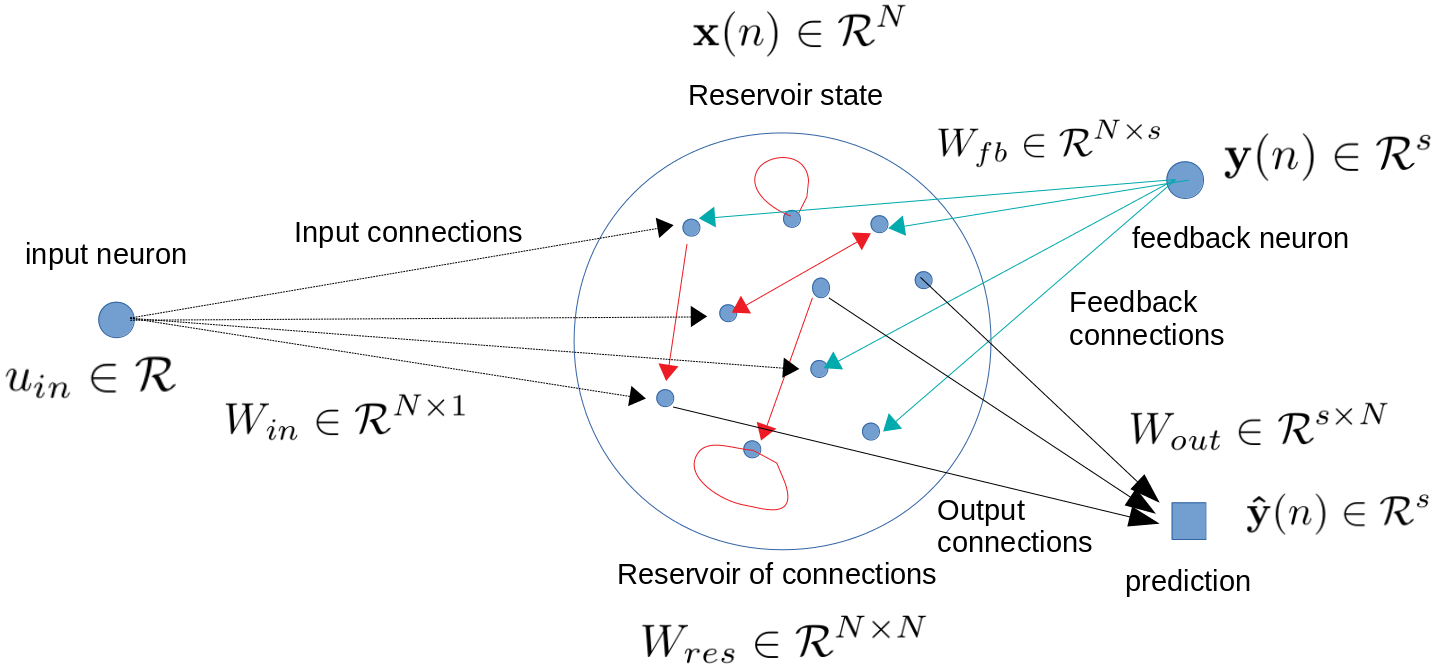}
 \end{center}
 \caption{\label{fig:reservoir}Reservoir computing: The reservoir
    is a collection of $N$ nodes. The state of the reservoir is given by the
    state vector $\xx$. The connections between nodes of the reservoir, $\Wres$,
    depicted by red arrows, are taken from a large, sparse, random matrix
    with a spectral radius less than unity.  During \textit{training},  the time-series, $\yy$,
    number of sunspots as a function of time, is fed into the feedback neuron. The update rule for each node is
    $ \xx(n+1) = \tanh\left( u_{\rm in}\Win + \Wres\xx(n) + \Wfb\yy \right)$,
    where $\Win$ and $\Wfb$ are random weights and $u_{\rm in}$ is a constant. 
    The output of the reservoir, $\yhat$, is
    a weighted sum -- with weights $\Wout$ --  over the state of the reservoir. 
    To forecast, the output of the reservoir is fed into the feedback neuron. 
}
\end{figure*}

Let us first briefly introduce the idea of reservoir computing as applied to the
problem of forecasting the next solar cycle, see \fig{fig:reservoir}. 
At the heart of the algorithm lies a neural network
with a large number of nodes -- the reservoir.
Every node of this network is called a neuron.
The state of the reservoir is given by the
state vector of dimension $N$, $\xx$.
Every neuron  gets its input from
all the neurons in the network (including itself),
a different input signal $u_{\rm in}$ (constant), and a feedback neuron. 
Each neuron is updated by operating a nonlinear
function (often tangent hyperbolic) on  a linear combination of
its inputs each multiplied by a different random weight:
\begin{equation}
  \xx(n+1) = \tanh\left( u_{\rm in}\Win + \Wres\xx(n) + \Wfb\yy \right)\/.
  \label{eq:update}
\end{equation}
The random weight that connects any two neurons of this network, $\Wres$, is given
by the corresponding element of  a large, sparse, random matrix
whose spectral radius is less than unity.
The linear combination of the output of each individual neuron
weighed by another set of weights, $\Wout$,
is the output of the reservoir:
\begin{equation}
  \yhat = \Wout \xx + \bb u_{\rm in}
 \label{eq:output}
\end{equation}
First, we must \textit{train} the reservoir. This proceeds in the following manner.
We treat the sunspot data -- number of sunspots as a function of time -- as
a time-series, $\yy$,  which is fed into the feedback neuron sequentially.
At every time instant an output of the reservoir is obtained.
The weights $\Wout$ are chosen to minimize a cost function -- the mean-square error
between the time-series and the output of reservoir;
\begin{equation}
  \mathcal{C} \equiv \mid \yy - \yhat \mid^2\/.
  \label{eq:cost}
  \end{equation}
All other weights in the algorithm are randomly chosen at the start and
are held constant.

A central feature of machine-learning techniques in general and reservoir computing
in particular is that to obtain a reliable forecast often very large amount of training
data is necessary.
The forecast is expected to get better longer the training period is
but there is no a-priori constraint on how long a training period is appropriate.
A related feature  -- bias-variance tradeoff~\citep{mehta2019high}
-- appears when the data available to train the algorithm
is limited. This is  true for almost any problem in natural sciences,
particularly so for the case of forecast of solar cycle.
If we optimize $\Wout$ such that the
output of the reservoir is a very good approximation to the training data
the forecast may actually become less reliable. 

To test how the algorithm performs when it is not constrained by limited data
we first apply this to a model of the solar dynamo. 
There are several low-dimensional, stochastic models that describe the same qualitative
features as the global sunspot data, v.i.z., oscillations whose frequency and
amplitude may change abruptly from one cycle to another. 
We use the model by \cite{hazra2014stochastically}. 
We run the dynamo code for a very long time.
We divide the time series into two parts. A very long training phase
and a short \textit{testing phase}, which is the last four cycles of the time series.
After the training phase we forecast the next four cycles and compare our results
against the test data. We obtain good agreement for the first cycle only. 
The details of these simulations will be reported elsewhere. 

Next we apply the algorithm to forecast solar cycles.
To be specific, let us consider the case of forecasting one particular cycle, say cycle 23.
We train the reservoir with the sunspot data with a thirteen-month
running average till the end of the cycle 22.
Then continue running to produce the forecast by feeding the output of the
reservoir to the feedback neuron. 
Note that every realization of the random matrix,
$\Wres$, gives a different forecast.
This presents us with a natural way to generate an ensemble of
forecasts by running our algorithm several times.
Our final forecast is the mean calculated over this ensemble.
\begin{figure*}
  \includegraphics[width=0.95\linewidth]{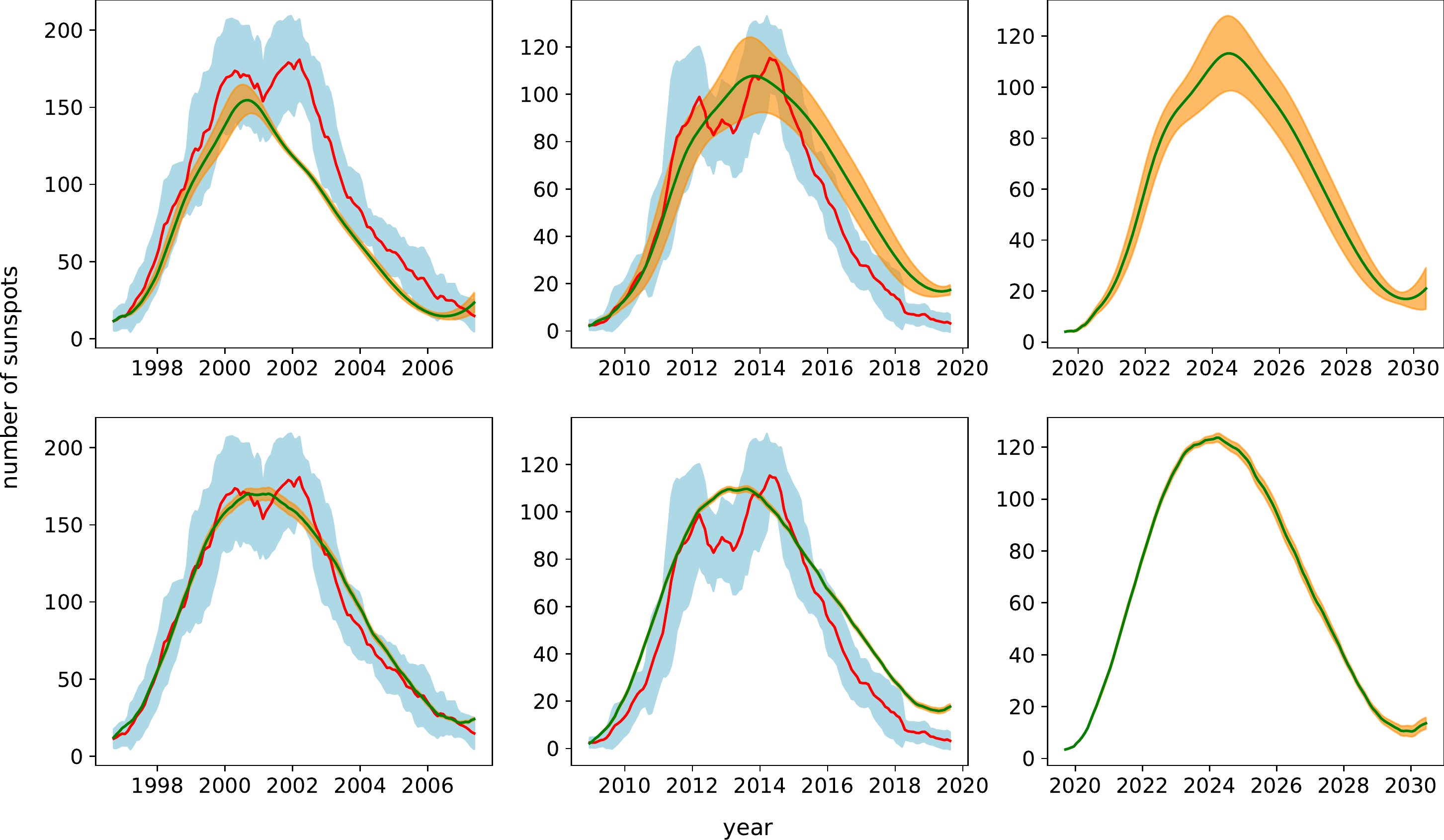}
  \caption{\label{fig:6forecast} Our forecasts for cycles 23, 24 and 25. We train the algorithm until the start of the cycle we want to forecast.
    For cycles 23 and 24 we compare our forecast with observation. The top
    panel shows our forecasts with the usual algorithm for reservoir computing.
    The red line shows the number of sunspots as a function of time with a thirteen-month moving window average. The shaded blue region shows the standard deviation of the observation. The green line is the forecast obtained by calculating the mean of the ensemble of forecasts. We use ten independent realizations. The orange shaded region shows the standard deviation of the ensemble. The bottom row shows the forecasts with a modified version of the
  standard algorithm. }
\end{figure*}

In the top panel of \fig{fig:6forecast} we show our forecasts for the cycles
23, 24 and 25. For the first two cycles we compare our forecast with
existing observations. For cycle 23 we find that the forecast is
quite accurate till the first peak. The standard-deviation (calculated
within the ensemble) increases with time
till the first peak but after the peak the standard-deviation decreases.
For cycle 24 the overall agreement is better,  the standard deviation steadily increases with time.

By its construction the reliability of the forecast decreases as time progresses, because
small errors at early times feeds into the algorithm and are magnified with time.
We develop a variation on the standard reservoir algorithm to improve the forecast
at late times. We make two changes. One, instead of feeding the input signal one
data point at a time, at time
$t$ we construct a $p$ dimensional vector $\yy(t-p+1), \yy(t-p+2),\ldots \yy(t) $.
This vector is fed to the feedback neuron. 
Two, we change the dimension of the output of the reservoir, such that we no
longer forecast one time instant after every update of the reservoir, but we
forecast a $q$ dimensional vector $\yhat(t+1) \ldots \yhat(t+q-1) $.
We use this to forecast one complete cycle in one go.
We adjust $p$ by trial and error to obtain the best result during the training phase.
The forecast from this algorithm is plotted in the bottom panel of \fig{fig:6forecast}.
Compared to the standard algorithm, our modified algorithm gives better results
when tested against the observation for both cycle 23 and 24.

Our forecast for the upcoming cycle 25 with both algorithms is shown in the
right most column of \fig{fig:6forecast}. Using the standard algorithm we
forecast that the maxima of the cycle 25 is going to appear between May and June of 2024
and that the maximum number of sunspots is going to be $113\pm 15$.
Our modified algorithm gives a maxima that is flatter, almost constant between
June 2023 and August 2024, with a maximum number of $124\pm2$ sunspots.
Note that the averaged sunspot data shows a distinctive two-peak behaviour 
in both cycles 23 and 24 -- this behavior is not present in all
sunspot cycles -- that is not captured by either of our algorithms.
We expect the same to happen for cycle 25 -- none of our
algorithms can forecast whether it may or
may not have this two-peak feature. Both our forecasts
show that the cycle 25 is expected to reach a minima 
near the beginning of the year 2030. 
Qualitatively cycle 25 is going to be weaker than cycle 23 but stronger than
cycle 24. 

Let us mention here few other recent forecasts of 
cycle 25. \cite{covas2019neural} forecast a weak cycle with a peak 
near 2022-2023. Their forecast for cycles 23 and 24 systematically 
overestimates the number of sunspots. According to \cite{pala2019forecasting} "the maximum in Solar Cycle 25 will be reached with peak SSN of 167.3 in July 2022 and Cycle 25 will last about ten years. This result means that Cycle 25 will be stronger than both Cycle 23 and Cycle 24.". They also provide a very useful summary of several earlier results. The forecast by \cite{benson2020forecast} shows that the upcoming Solar Cycle 25 will have a maximum sunspot number around $106 \pm 19.75$ 
-- it is going to be slightly weaker than cycle 24.

\begin{figure*}
    \includegraphics[width=\linewidth]{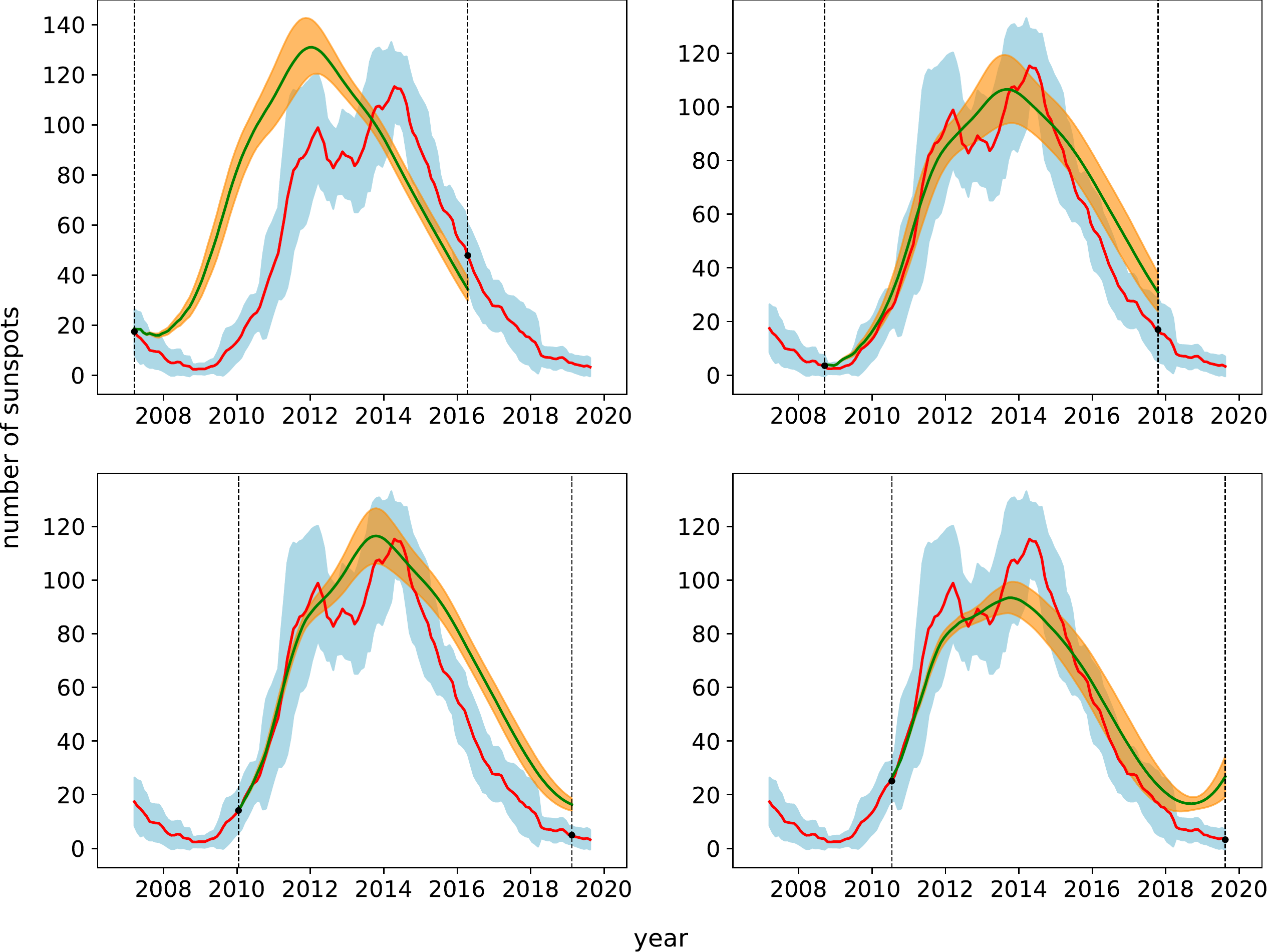}
  \caption{\label{fig:stop} The forecasts depend crucially on
    when we stop training and start forecasting. We show four
    representative cases with the standard algorithm for the
    cycle 24. 
  In each plot, a pair of dashed lines indicates the beginning and end of the forecast. 
  The best forecast (bottom right plot) is obtained when we stop training at 
   the rising phase of the cycle. } 
\end{figure*}

Next we point out several additional features that have emerged
from our work. First, our forecast beyond one cycle is inaccurate.
Machine--learning algorithms typically improve with more data, hence
we expect that  our methods applied to proxies of sunspot 
data may be more successful in forecasting beyond one cycle.
Second, our new method performs fairly well for the last 5 solar cycles
except cycle 21.
Finally we show that the reliability of the forecast depends crucially on when
we stop training and start forecasting.
This is what we expect naively. Near the minima of the cycle
the signal to noise ratio in the sunspot data is the lowest. 
Hence if we stop training at the lowest point of a cycle
we expect the worst result if the level of noise is roughly constant
as a function of time. A look at the sunspot data shows that
the noise is not a constant but is actually significantly large
near the peak of the cycle -- the noise increases as the signal increases.
Nevertheless the signal to noise ratio is less than unity near the
minima of the cycles. 
To check this expectation,  
in \fig{fig:stop} we  show four representative cases with the
standard algorithm for the cycle 24.
 The best forecast is obtained when we are in
the rising phase of the cycle.
At present we are very near the minima of a solar cycle hence we
expect our forecast to improve if we recalculate our results
approximately a year later.
The performance of the  standard reservoir algorithm depends much more
sensitively on the exact time when training is stopped compared to
our new algorithm -- the standard algorithm  becomes unstable
if the training is stopped very close to the present minima. 

To conclude, we use two different algorithms, one standard reservoir computing
and the other one a modification of it, to forecast forthcoming cycle 25.
Both algorithms agree that cycle 25 is going to last for about 10 years.
The maxima is going to be reached in the year 2024. As for the maximum
number of sunspots, the standard algorithm forecasts it to be 113 ($\pm15$) 
whereas our novel method forecasts 124 ($\pm2$). 

\acknowledgments
The collaboration between NORDITA and KTH on the one hand and IISER Kolkata on the other hand
has been made possible through the SPARC project grant SPARC/2018-2019/P746/SL of the Government of India. The Center of Excellence in Space Sciences India is funded by the Ministry of Human Resource Development, Government of India under the Frontier Areas of Science and Technology (FAST) scheme. DM is supported by  two Swedish research council grants, Nos. 638-2013-9243 and 2016-05225.
DM thanks the participants in the Machine-Learning seminars at NORDITA
organized by S-H Lim for useful inputs. 
AEF's stay at NORDITA is supported by a grant from the Swedish research council.  
Part of the work was done during an earlier visit by AEF to Nordita. The visit was supported by the Erasmus+ Programme, where AEF did a three months internship at NORDITA as a recent graduate from the Univeristy of Barcelona. DN acknowledges a Visiting Professorship grant from the Swedish Wenner-Gren Foundation for facilitating his visit to NORDITA, Stockholm in 2018-- when ideas related to this project were developed.    

\bibliography{sunref}{}
\bibliographystyle{aasjournal}

\end{document}